\documentclass[prd,nofootinbib,floats,superscriptaddress,eqsecnum,tightenlines,11pt]{revtex4}

\usepackage{hyperref}
\usepackage{graphicx}
\usepackage{amsmath,amssymb,amsfonts,amsthm,latexsym,stmaryrd}
\usepackage{marginnote}
\usepackage{color}

\def\beq{\begin{equation}}
\def\eeq{\end{equation}}
\newcommand{\bea}{\begin{eqnarray}}
\newcommand{\eea}{\end{eqnarray}}
\def\bi{\begin{itemize}}
\def\ei{\end{itemize}}
\def\ba{\begin{array}}
\def\ea{\end{array}}
\def\bfig{\begin{figure}}
\def\efig{\end{figure}}

\def\tA{A} 
\def\B{{\cal B}}
\def\C{{\cal C}}
\def\a{\alpha}
\def\f{f}
\newcommand\h{\gamma}

\def\aa{\alpha_2}
\def\ab{\alpha_1}
\def\ac{\alpha_3}
\def\ad{\alpha_4}
\def\ae{\alpha_5}

\def\ba{\beta_1}
\def\bb{\beta_2}

\def\ka{\kappa_2}
\def\kb{\kappa_1}
\def\kc{\kappa_3}
\def\kd{\kappa_4}
\def\ke{\kappa_5}

\def\An{A_*}
\def\A{{\cal A}}

\def\U{{\cal U}}
\def\R{{}^{(4)}\!R}

\def\fX{f_{,X}}
\def\fphi{f_{,\phi}}

\newcommand{\Gfour}{G_4{}}
\newcommand{\Ffour}{F_4{}}

\begin{document}

\title{Hamiltonian analysis of higher derivative scalar-tensor theories}
\author{David  Langlois}
\email{langlois@apc.univ-paris7.fr}
\affiliation{Laboratoire APC -- Astroparticule et Cosmologie, Universit\'e Paris Diderot Paris 7, 75013 Paris, France}
\author{Karim Noui}
\email{karim.noui@lmpt.univ-tours.fr}
\affiliation{Laboratoire de Math\'ematiques et Physique Th\'eorique, Universit\'e Fran\c cois Rabelais, Parc de Grandmont, 37200 Tours, France}
\affiliation{Laboratoire APC -- Astroparticule et Cosmologie, Universit\'e Paris Diderot Paris 7, 75013 Paris, France}

\begin{abstract}
We perform  a  Hamiltonian analysis of a  large class of scalar-tensor Lagrangians which depend quadratically on the  second derivatives of a scalar field.  By resorting to a convenient choice of dynamical variables, we show that the Hamiltonian can be written in a very simple form, where  the Hamiltonian and the momentum constraints are easily identified. In the case of degenerate Lagrangians, which include the Horndeski and beyond Horndeski quartic Lagrangians, our analysis confirms that the dimension of the physical phase space is reduced by the primary and secondary constraints due to the degeneracy, thus leading to the elimination of the dangerous Ostrogradski ghost. We also present the Hamiltonian formulation for nondegenerate theories  and find that they contain four degrees of freedom, as expected. We finally discuss the status of the unitary gauge from the Hamiltonian perspective. 
\end{abstract}

\maketitle

\section{Introduction}
As a possible alternative explanation for the observed cosmological acceleration, theories of modified gravity have attracted considerable interest in the recent past  (see e.g. \cite{Clifton:2011jh,Joyce:2014kja,Berti:2015itd,Koyama:2015vza} for reviews). Many, although not all, models of modified gravity are based on scalar-tensor theories, where one scalar degree of freedom is  combined with the gravitational metric. The models that have been studied in the literature have progressively increased in complexity and generality, from quintessence models up to Lagrangians involving  second-order derivatives of the scalar field. In the latter case, special care must be taken to avoid the so-called Ostrogradski instability~\cite{Ostrogradsky}. Indeed, second or higher order time derivatives in the Lagrangian generically lead to the presence of  an extra degree  of freedom, which  behaves like a ghost. It has long been believed that, in order to avoid Ostrogradski's ghost, it was necessary that the Lagrangian yields second-order equations of motion.  This property is indeed satisfied  for  flat spacetime galileon~\cite{NRT}. When gravity is included, the same requirement led  to the so-called generalized galileons~\cite{Deffayet:2009mn,Deffayet:2011gz}, which   coincide with Horndeski theories~\cite{Horndeski:1974wa,Kobayashi:2011nu} in four dimensions. 

The statement that second-order equations of motion are necessary to avoid ghost-like instabilities was first questioned in \cite{Zumalacarregui:2013pma} by considering a theory obtained from Einstein-Hilbert via disformal transformation, and  in \cite{Gleyzes:2014dya}, in the context of Horndeski theories, by proposing two extensions, denoted $L_4^{\rm bH}$ and $L_5^{\rm bH}$,  of the quartic and quintic Horndeski Lagrangians $L_4^{\rm H}$ and $L_5^{\rm H}$ (further extensions, leading to Lorentz-breaking theories,  were also proposed in \cite{Gao:2014soa}). 
It was later demonstrated in \cite{Gleyzes:2014qga} that combinations of $L_4^{\rm H}$ and $L_4^{\rm bH}$, on one hand, or combinations of $L_5^H$ and $L_5^{bH}$, on the other hand, can be related to a purely Horndeski Lagrangian via a disformal transformation, thus indicating that the number of degrees of freedom in these subclasses beyond Horndeski should be the same as in Horndeski theories.
This brought  further confirmation that at least some combinations of  terms beyond Horndeski  with Horndeski's ones were indeed healthy, although the status of arbitrary  combinations of all terms  remained uncertain.

Further support was apparently provided by a recent work \cite{Deffayet:2015qwa} in which it was shown that   the third-order covariant  equations of motion for the scalar field and the metric can be rewritten,  in an arbitrary gauge, as a system of equations which are second-order in time derivatives. However, it is not fully clear what this implies about the number of degrees of freedom,  as noted by the authors themselves.  The same paper also presented  a Hamiltonian treatment, in an arbitrary gauge, of the particular Lagrangian $L_4^{\rm bH}$  and showed that the number of degrees of freedom is strictly less than four. 

In a previous paper~\cite{Langlois:2015cwa}, we reconsidered the question of the Ostrogradski ghost in higher derivative scalar-tensor theories from a  different perspective, by focusing on the degeneracy of the Lagrangian. As we showed, the notion of degeneracy is much richer when a variable with second-order time derivatives is coupled to other degrees of freedom than when it is isolated.  Working in an arbitrary coordinate system, we explored the degeneracy of a large class of scalar-tensor theories for which the Lagrangian depends quadratically on second derivatives of the scalar field. This class includes 
$L_4^{\rm H}$ and $L_4^{\rm bH}$ and we could thus demonstrate that these two Lagrangians are degenerate, as well as their sum. We also found other degenerate Lagrangians, which do not belong to the extensions beyond Horndeski introduced in \cite{Gleyzes:2014dya}. 

Furthermore, we investigated the special case of $L_5^{\rm bH}$, which is cubic in second derivatives of the scalar field, and found that it is also degenerate, as well as the combination $L_4^{\rm bH}+L_5^{\rm bH}$. By contrast, we noticed that the combination $L_4^{\rm H}+L_4^{\rm bH}+L_5^{\rm bH}$ is {\it not} degenerate, which suggests that combinations of  Horndeski terms with both quartic and quintic terms beyond Horndeski are not viable in general. Note that this result is  compatible with the conclusions  of \cite{Gleyzes:2014qga} concerning disformal transformations since the above combination {\it cannot} be related to Horndeski via disformal transformation. 
However, it may seem at odds with the unitary gauge Hamiltonian analysis of \cite{Gleyzes:2014dya,Lin:2014jga,Gleyzes:2014qga}, or rather its extrapolation for the quintic terms (as the detailed analysis was in fact restricted to the quartic terms). 
This apparent paradox is resolved by the fact that some nondegenerate, and thus unhealthy, theories can appear degenerate in the unitary gauge, as  discussed in~\cite{Langlois:2015cwa}. 
As the unitary gauge can sometimes be misleading, it is worth revisiting the Hamiltonian analysis of higher derivative theories by considering an arbitrary gauge and check whether we can confirm our conjecture that healthy theories, i.e. without Ostrogradski ghost, correspond to degenerate theories. 

In this  work, we  present the full Hamiltonian analysis for the class of models studied in our previous work. By using appropriate dynamical variables, we are able to write the full Lagrangian in a relatively compact form, which greatly simplifies the computation of the Hamiltonian. 
The structure of the Hamiltonian is rather simple and exhibits, like in general relativity, a term linear in the lapse function and another linear in the shift. One thus recognizes the  structure associated with spacetime diffeomorphisms invariance. This enables us to identify the first-class constraints generating the diffeomorphism invariance.
For the detailed Hamiltonian analysis, one needs to distinguish between degenerate theories and nondegenerate ones. In the former case, the computation is a bit more involved because of the presence of a primary constraint between momenta, which in turn generates a secondary constraint.  These two constraints, which are second-class, eliminate one degree of freedom in comparison with nondegenerate theories. One thus ends up with three degrees of freedom for degenerate theories, compared with the four degrees of freedom for nondegenerate theories.

This paper is organized as followed. In section 2, we present our general action and compute the full ADM decomposition of the action in an arbitrary gauge. In section 3, we focus on the kinetic terms of the Lagrangian, which we rewrite as a bilinear form acting on a 7-dimensional vector space. Section 4, which is the main section of this paper, is devoted to the Hamiltonian formulation of degenerate theories, the identification of first-class and second-class constraints and the counting of the number of degrees of freedom. In section 5, we repeat the same analysis for nondegenerate theories. We then discuss the 
unitary gauge in Section 6. We summarize our results in the final section. Some technical details are given in three Appendices.

\section{General action and $3+1$ ADM decomposition}
In this section, we perform the (3+1)-decomposition of the action, which is a prerequisite for the Hamiltonian analysis. We consider  scalar-tensor actions of the form
\beq\label{generalHorn}
S[g_{\mu\nu},\phi] \equiv \int d^4x \, \sqrt{\vert g \vert} \left[ \f \, {\cal R} + C^{\mu\nu\rho\sigma}\,  (\nabla_\mu \nabla_\nu\phi) \, (\nabla_\rho \nabla_\sigma\phi) \right]
\eeq
where ${\cal R}$ is the 4-dimensional Ricci scalar, and  the tensor $C^{\mu\nu\rho\sigma}$, which depends only on $\phi$ and $\phi_\mu\equiv \nabla_\mu\phi$, can always be written as
\begin{eqnarray}\label{family}
C^{\mu\nu\rho\sigma} & \equiv &  \frac{1}{2} \ab\, (g^{\mu\rho} g^{\nu\sigma} + g^{\mu\sigma} g^{\nu\rho})+\aa \,g^{\mu\nu} g^{\rho\sigma} +\frac{1}{2} \ac\, (\phi^\mu\phi^\nu g^{\rho\sigma} +\phi^\rho\phi^\sigma g^{\mu\nu} ) 
\cr
& + &   \frac{1}{4} \ad (\phi^\mu \phi^\rho g^{\nu\sigma} + \phi^\nu \phi^\rho g^{\mu\sigma} + \phi^\mu \phi^\sigma g^{\nu\rho} + \phi^\nu \phi^\sigma g^{\mu\rho} ) +  \ae\, \phi^\mu \phi^\nu \phi^\rho \phi^\sigma \label{four} .
\end{eqnarray}
Here $f$ and $\a_i$ are functions of $\phi$ and $X=\phi_\mu \phi^\mu$ only. 

\subsection{Particular cases}
The class of theories (\ref{generalHorn}) includes as a particular case the quartic Horndeski term
\beq
L^{\rm H}_4 = \Gfour(\phi,X) \, \R - 2 \Gfour_{,X}(\phi,X) (\Box \phi^2 - \phi^{ \mu \nu} \phi_{ \mu \nu}) \,.
\eeq
The above Lagrangian is indeed of the form (\ref{generalHorn})-(\ref{family}) with 
\beq
f=\Gfour\,, \qquad \ab= -\aa= 2 \Gfour_{,X}\,, \qquad \ac=\ad=\ae=0\,.
\eeq
The action (\ref{generalHorn}) also includes the extension beyond Horndeski introduced in \cite{Gleyzes:2014dya}, which can be written as
\beq
L^{bH}_4=\Ffour(\phi,X) \epsilon^{\mu\nu\rho}_{\ \ \ \ \sigma}\, \epsilon^{\mu'\nu'\rho'\sigma}\phi_{\mu}\phi_{\mu'}\phi_{\nu\nu'}\phi_{\rho\rho'}\,.
\eeq
This corresponds to (\ref{generalHorn})-(\ref{family}) with 
\beq
\ab=-\aa= X \Ffour     \,, \qquad \ac=-\ad= 2 \Ffour\,, \qquad \ae=0\,.
\eeq
Various aspects of these theories beyond Horndeski have been investigated recently (see e.g~\cite{Kase:2014yya,Fasiello:2014aqa,Kobayashi:2014ida,Bettoni:2015wta,Koyama:2015oma,Saito:2015fza,DeFelice:2015isa,Tsujikawa:2015mga,Lombriser:2015cla,Tsujikawa:2015upa,Babichev:2015qma,Sakstein:2015zoa,Kase:2015zva,Akita:2015mho}). 

\subsection{ADM decomposition and notations}
In the ADM formalism,  the metric  $ds^2=g_{\mu\nu} dx^\mu dx^\nu$ is parametrized as follows 
\bea
ds^2&=&-N^2 dt^2 +\h_{ij} (dx^i+N^i dt)(dx^j+N^j dt)\,,
\label{metric_ADM}
\eea
where $N$  is the lapse function and $N^i$ the shift vector. In matricial form, the metric $g_{\mu\nu}$ and its inverse $g^{\mu\nu}$ are given by
\beq
g_{\mu\nu}=
\left(
\begin{array}{cc}
-N^2+\h_{ij}N^i N^j & \h_{ij}N^j
\\
\h_{ij}N^i & \h_{ij}
\end{array}
\right)\,
\;\;\text{and}\;\;
g^{\mu\nu}= \frac{1}{N^2}
\left(
\begin{array}{cc}
-{1}& {N^j}
\\
{N^i} & N^2\h^{ij}-{N^i N^j }
\end{array}
\right)\,.
\eeq
We also need to introduce the second fundamental form which is  given by 
\beq
K_{ij}=\frac{1}{2N}\left( \dot{\h}_{ij}-D_iN_{j}-D_jN_{i}\right)\,,
\label{Eij}
\eeq
where  $D_i$ denotes the spatial covariant derivative associated with 
the spatial metric $\h_{ij}$.

In order to make the ADM decomposition of the action, it is convenient to replace second order  derivatives that appear in (\ref{generalHorn}) by first order derivatives via the introduction of new dynamical variables. 
We thus consider the new action
\begin{eqnarray}\label{modifiedHorn}
S[g_{\mu\nu},\phi;A_\mu,\lambda^\mu] \equiv \int  d^4x \,  \sqrt{\vert g \vert} \left(\f\,  {\cal R} + C^{\mu\nu\rho\sigma} \nabla_\mu A_\nu \, \nabla_\rho A_\sigma \right) + \lambda^\mu (\nabla_\mu\phi - A_\mu),
\end{eqnarray}
which contains the auxiliary field $A_\mu$,  as well as the vector field $\lambda^\mu$ enforcing the relation 
\begin{eqnarray}\label{A=phimu}
A_\mu=\nabla_\mu \phi \,.
\end{eqnarray}
Note that for $\mu=0$, the previous relation is an equation of motion whereas it is a constraint for $\mu=i$. 
It is easy to show that this action is indeed equivalent to
the original one (\ref{generalHorn}) when one writes the Euler-Lagrange equations.
In this new formulation,  the tensor $C^{\mu\nu\rho\sigma}$  depends $A_\mu$ (and no longer on $\phi_\mu$).

Furthermore, as a consequence of  (\ref{A=phimu}), $A_\mu$ satisfies the symmetry relation
$\nabla_\mu A_\nu = \nabla_\nu A_\mu$. When distinguishing temporal and spatial indices, this property allows us to replace  all the terms  $\nabla_0 A_i$ by $\nabla_i A_0$ in the action without changing the equations of motion, as shown explicitly in Appendix \ref{Appendix_A}.

\subsection{Einstein-Hilbert term}
 We first present the ADM decomposition of the  Einstein-Hilbert Lagrangian multiplied by a function of $\phi$ and $X=g^{\mu\nu} A_\mu A_\nu$, corresponding to the action
 \bea
S_{EH} = \int d^4x \, \sqrt{\vert g \vert} \, f \, {\cal R}\,.
\eea
As is well-known, the (3+1) decomposition of this  action yields
\beq
S_{EH}= \int dt \, d^3 x \, N\sqrt{\h}\, f \, \left(K_{ij} K^{ij}-K^2+R - 2 \nabla_\mu (a^\mu - K n^\mu)\right)\,,
\label{action-ADM}
\eeq
where  $\h\equiv$ det$(\h_{ij})$, $R$ is the 3-dimensional Ricci scalar, $K_{ij}$ is the second fundamental form   (\ref{Eij}) and $K=K_i^i$ is its trace.
The last term in the action (\ref{action-ADM}) involves  the acceleration $a^\mu$ and the normal $n^\mu$ (of the spatial hypersurface $\Sigma$ )  whose  components are
\beq
 a^\mu=n_\nu \nabla^\nu n^\mu \; ,\qquad  \; n^\mu = \frac{1}{N} (1,-N^i) \,.
\eeq
When $f$ is constant, the last term in the action is a total derivative, which can be discarded. This term however becomes relevant when $f$ depends on the scalar field or its derivatives. 
To perform the (3+1)-decomposition of this term, it is convenient to introduce the new variable
\begin{eqnarray}
\label{def_An}
\An \equiv A_\mu n^\mu = \frac{1}{N}(A_0 - N^i A_i) \,,
\end{eqnarray}
which corresponds  to the normal  component of $A_\mu$ with respect to the spatial hypersurface $\Sigma$.  

After a straightforward calculation, we find
\begin{eqnarray}
S_{EH} = \int dt \, d^3 x \, N\sqrt{\h}\, \left( \frac2N {\cal B}_{\rm grav}^{ij} K_{ij} (\dot\An-\Xi)  + {\cal K}^{ij,kl}_{\rm grav} K_{ij} K_{kl} + 2 {\cal C}_{\rm grav}^{ij} K_{ij} - \U_{\rm grav}\right)
\end{eqnarray}
where we have introduced the function
\beq
\Xi\equiv \tA^k D_k N + N^k D_k\An\,,
\eeq
and where the coefficients entering in the Lagrangian are given by
\begin{eqnarray}
 {\cal B}_{\rm grav}^{ij} & = & 2 \fX \An \h^{ij} \, ,\\
 {\cal K}^{ij,kl}_{\rm grav}  & = & \frac{1}{2} f ( \h^{ik} \h^{jl} + \h^{il} \h^{jk} - 2 \h^{ij} \h^{kl} )+ 
 2 \fX ( \h^{ij} \tA^k \tA^l + \h^{kl} \tA^i \tA^j ) \, ,\\
 {\cal C}_{grav}^{ij} & \equiv &  - \h^{ij}  ( 2\fX \tA^k (D_k\An) +  \fphi  \An)  
   \, ,\\
 \U_{\rm grav} & = & -R + 2 D_i (\fX D^i X + \fphi \tA^i) \; .
\end{eqnarray}
We use the notations $\fphi \equiv \partial f/\partial \phi$ and $\fX\equiv \partial f / \partial_X$ for partial derivatives. All spatial indices are raised or lowered by the spatial metric $\h_{ij}$. In particular, we define $A^i\equiv \h^{ij}A_j$, so that  $X=- \An^2 + A_i \tA^i$.

\subsection{Scalar-tensor interaction term}

We proceed in a similar way to decompose the ``scalar-tensor" interaction part of the action
\begin{eqnarray}\label{scalartensorcoupling}
S_\phi \equiv \int d^4x \, \sqrt{\vert g \vert}  C^{\mu\nu\rho\sigma} \nabla_\mu A_\nu \, \nabla_\rho A_\sigma \,.
\end{eqnarray} 
In that case, we need to compute the components of the tensor 
\bea
A_{\mu\nu}\equiv \nabla_\mu A_\nu \equiv \partial_\mu A_\nu - \Gamma_{\mu \nu}^\rho \, A_\rho\,.
\eea
Using the expressions of the  Christoffel symbols $\Gamma_{\mu\nu}^\rho$ in term of ADM quantities, given in Appendix \ref{Christoffel appendix},  one can easily obtain the different components of the covariant derivative of $A_\mu$
\begin{eqnarray}
A_{00}&=&N\dot\An -\left(\An  N^i N^j+2N A^{(i} N^{j)}\right)K_{ij} +N N^kD_k\An +N^iN^jD_iA_j
\cr
&&\qquad -NA^kD_kN +N^k(\dot A_k-D_k A_0)  \; ,\label{nabla000}
\\
A_{i0} &=&
-(\An  N^j+NA^j)K_{ij} +ND_i\An +N^kD_iA_k   \; ,
\\
A_{0i}&=& (\dot A_i-D_i A_0)
-(\An  N^j+NA^j)K_{ij}+ND_i\An +N^kD_iA_k  \; , \label{nabla00i}
\\
A_{ij}&=&  D_i A_j-\An  K_{ij}\,.
\end{eqnarray}
As discussed in Appendix  \ref{Appendix_A},  
the terms $(\dot A_k-D_kA_0 )$ and $(\dot A_i-D_i A_0)$, 
which appear in (\ref{nabla000}) and (\ref{nabla00i}), can be eliminated.  In this way, all the time derivatives of $A_i$ disappear from the action.

Using the results of previous subsections and  after a long calculation, 
one finds that the ADM decomposition of $S_\phi$ reduces to the following form
\begin{eqnarray}
S_\phi =\int N \sqrt{\h} \left[ \frac{{\cal A}}{N^2} \, (\dot\An-\Xi)^2 + \frac2N {\cal B}_\phi^{ij} (\dot\An-\Xi) K_{ij} + {\cal K}_\phi^{ijkl} K_{ij} K_{kl} +2 {\cal C}_\phi^{ij} K_{ij} +2 \frac{{\cal C}^0}{N} (\dot\An-\Xi) - \U_\phi\right]\quad \,.
\end{eqnarray}
We have not labelled  $\cal A$ and ${\cal C}_0$ with the subscript $\phi$ because such terms do not show up in the Einstein-Hilbert part of the action.
The coefficients of the quadratic terms in time derivatives have already been computed in \cite{Langlois:2015cwa}, and are given by\footnote{Our definitions for ${\cal A}$ and ${\cal B}^{ij}$ in  \cite{Langlois:2015cwa}  differ from the present ones by factors of the lapse $N$.}
\bea
\A & = & \aa+\ab-(\ac+\ad)\An^2+ \ae \An^4\,, \\
{\cal B}_\phi^{ij} & = &\frac{\An }{2}\left(2\aa-\ac\An^2\right)  \h^{ij} -\frac{\An }{2}\left(\ac+2\ad-2\ae\An^2\right) \,  \tA^i\tA^j \,, \\
{\cal K}_\phi^{ij,kl} &=&\ab \An^2 \h^{i(k}\h^{l)j} +\aa \An^2 \h^{ij} \h^{kl}- \frac12 \ac\An^2\left(\tA^i\tA^j \h^{kl}+\tA^k \tA^l \h^{ij}\right)
    \cr
  &&
  -\ab  \left(\tA^i\tA^{(k}\h^{l) j}+\tA^j\tA^{(k} \h^{l)i}\right) 
  +(\ae\An^2-\ad) \tA^i \tA^j\tA^k\tA^l\,.
\eea
The coefficients of the linear terms are
\bea
{\cal C}^{ij}_{\phi}  &=& \An (D_k A_l) [ -\ab \h^{ik} \h^{jl} -\aa \h^{kl}\h^{ij}+ \frac12\ac (\h^{kl} \tA^i\tA^j -\tA^k \tA^l \h^{ij}) + \ae \tA^i\tA^j\tA^k\tA^l]
  \cr
 &&+(D_k \An) [\tA^k ((\ad - 2\ae \An^2) \tA^i\tA^j + \ac \An^2 \h^{ij}) + \ab (\h^{jk} \tA^i + \h^{ik} \tA^j)] \, ,\nonumber \\
  {\cal C}^0 & = & \frac12(D_i A_j)[(2\ae \An^2-\ac)\tA^i\tA^j + (\ac \An^2 - 2\aa)\h^{ij}] + (\ac + \ad-2\ae \An^2) \An \tA^i D_i \An\,.
\eea
Finally, the potential is given by 
\bea
\U_\phi & = &- (\ab \h^{ik}\h^{jl} + \aa \h^{ij} \h^{kl} + \ac \tA^i \tA^j \h^{kl}+\ad \tA^i\tA^k \h^{jl} + \ae \tA^i  \tA^j  \tA^k  \tA^l) (D_i A_j)(D_k A_l) 
\cr
&&-  (4\ae \An^2  - \ad) \tA^i \tA^j (D_i \An)(D_j \An)- (\ad \An^2 - 2\ab) (D_i \An)(D^i \An) 
\cr
&&+2 \An  \left(2\ae \tA^i\tA^j\tA^k D_j A_k + \ac\tA^i D_j\tA^j+\ad \tA^j D_j\tA^i \right) D_i \An \,.
\eea
It is worth noticing that by using $\An$ instead of $A_0$, we have automatically absorbed time derivatives of the lapse and of the shift, which otherwise would appear explicitly in the action. In general, terms that depend on  $\dot N$ or $\dot N^i$ indicate the presence of additional degrees of freedom, but this is not always the case, as illustrated explicitly, for disformal transformations,  in \cite{Gleyzes:2014rba} and discussed  in more detail in \cite{Domenech:2015tca}.

\subsection{Summary: full (3+1) decomposition of the action}
Putting together all our previous results, we finally obtain  the full (3+1) decomposition of the modified action (\ref{modifiedHorn}):
\bea\label{3+1decomp}
S  & = & \int  dt \, d^3x \, N \sqrt{\h} \left[{\cal A} \, V_*^2 +
2 {\cal B}^{ij} V_* K_{ij} + {\cal K}^{ijkl} K_{ij} K_{kl} +2 {\cal C}^{ij} K_{ij} +2 {\cal C}^0 V_* - \U\right]
 \cr
  & + & \int dt \, d^3x \, ( p_\phi \dot\phi - N p_\phi \An - N^i p_\phi A_i + \lambda^i (\phi_i - A_i))
\eea 
where we have introduced the quantity
\beq
V_*\equiv \frac1N (\dot\An-\Xi)\,.
\eeq
The  coefficients are given by 
\bea
{\cal B}^{ij}={\cal B}_{\rm grav}^{ij}+
{\cal B}_\phi^{ij} \;\; , \;\;
 {\cal K}^{ijkl} =  {\cal K}_{\rm grav}^{ijkl} +  {\cal K}_\phi^{ijkl} \;\;, \;\;
 {\cal C}^{ij} = {\cal C}^{ij}_{\rm grav}  + {\cal C}_\phi^{ij}  \;\; , \;\;
 \U=\U_{\rm grav} + \U_\phi \,.
\eea
In particular, the tensorial structure of the coefficients ${\cal B}^{ij}$ and ${\cal K}^{ijkl}$, which appear in the kinetic part of the action, depends only on the metric $\h_{ij}$ and the vector $A_i$:
\bea
{\cal B}^{ij} & = & \ba \h^{ij} + \bb \tA^i \tA^j \\
{\cal K}^{ij,kl} & = &  \kb \h^{i(k}\h^{l)j} +\ka\, \h^{ij} \h^{kl}+  \frac12 \kc\left(\tA^i\tA^j \h^{kl}+\tA^k \tA^l \h^{ij}\right)
    \cr
  &&
  +\frac 12 \kd  \left(\tA^i\tA^{(k}\h^{l) j}+\tA^j\tA^{(k} \h^{l)i}\right) 
  +\ke \tA^i \tA^j\tA^k\tA^l\,,
\eea
 with 
 \begin{eqnarray}
&& \ba=\frac{\An}{2}(2\aa - \ac \An^2 + 4\fX) \,, \quad \bb=\frac{\An}{2} (2\ae \An^2 - \ac - 2\ad)  \\
&&\kb=\ab \An^2 + f\,, \, \ka=  \aa \An^2-f\,, \,    \kc=- \ac\An^2+4\fX\,, \,  \kd=- 2\ab\,,\,  \ke=\ae\An^2-\ad\,.
 \end{eqnarray}
In anticipation of the Hamiltonian analysis, we have changed the notation $\lambda^0$ into $p_\phi$.
Note that the coefficients $\cal A$ and ${\cal C}^{0}$ are unaffected by the gravitational part of the action.

\section{Kinetic terms and degeneracy condition}
The kinetic part of  the action is given  by the expression
\bea\label{kineticnewbasis}
S_{\rm kin} =  \int dt \, d^3x \, N \sqrt{\h} \, {\cal L}_{kin}  \;�\; \text{with} \;\;  {\cal L}_{\rm kin}  = {\cal A} \, V_*^2 +
2
  {\cal B}^{ij} V_* K_{ij} + {\cal K}^{ij,kl} K_{ij} K_{kl} \,, 
\eea
where ${\cal L}_{\rm kin}$ can be viewed as a  bilinear form acting on a 7-dimensional vector space (the vector space of $3\times 3$ symmetric matrices is 6-dimensional). For a better understanding of the structure of the kinetic terms, it is instructive to introduce a basis where ${\cal L}_{\rm kin}$ can be diagonalized, or at least  block diagonalized.

\subsection{Metric kinetic terms}

Let us first concentrate on  ${\cal K}^{ij,kl}$ which defines a  bilinear form
on the 6-dimensional space of $(3\times 3)$ symmetric real matrices $\text{Sym}(3)$, or, equivalently, a linear map
\bea
{\cal K}: \text{Sym}(3) \longrightarrow \text{Sym}(3) \;\;\; , \;\;\; U \longmapsto {\cal K} U \;\; \text{s.t.} \;\; ({\cal K} U)^{ij} = {\cal K}^{ij,kl} U_{kl} \,.
\eea
The space $\text{Sym}(3)$ is naturally endowed with  the scalar product
\bea
\langle U , V \rangle =  U_{ij} \h^{jk}  V_{kl} \h^{li}=U_{ij}V^{ij}\,,
\eea
and one can try to construct an  orthonormal basis of  $\text{Sym}(3)$, with respect to this scalar product, in which  ${\cal K}$ takes a simple form. To do so, let us introduce  two unit  spatial vectors $u^i$ and $v^j$ so that they form, together with the normalized vector $A^i/\|A\|$ (where $\|A\|=\sqrt{\tA^2}$), a complete orthonormal basis in 3-dimensional space, i.e. such that
\bea
u^i u_i = v^i v_i = 1 \;\; , \;\; u^i v_i = v^i A_i = A^i u_i = 0 \, .
\eea 
By using these vectors, one can  build an orthonormal basis of $\text{Sym}(3)$, which consists of   the following independent 6 matrices $U^I$:
\bea
&&U^1_{ij} = \frac{1}{\|A\|^2} A_i A_j  \;\;\; , \;\;\; U^2_{ij} = \frac{1}{\sqrt{2}} (\h_{ij} - U^1_{ij}) \;\;\; , \;\;\; U^3_{ij} = \frac{1}{\sqrt{2}}(u_iu_j - v_i v_j)\,, \cr
&&U^4_{ij} = \frac{1}{\sqrt{2}}(u_i v_j + u_j v_i) \; , \;\;\; U^5_{ij} = \frac{1}{ \sqrt{2} \|A\| }(u_i A_j + u_j A_i) \; , \;\;\; U^6_{ij} = \frac{1}{\sqrt{2}  \|A\|}(v_i A_j + v_j A_i) \,.
\eea

An immediate calculation shows that ${\cal K}$ is block diagonal in this basis. Indeed the four vectors $U^I$ for $I \in \{ 3,4,5,6\}$ are eigenvectors of 
${\cal K}$, while  the subspace spanned by $(U^1, U^2)$ is stable under the action of ${\cal K}$. More precisely, we have
\bea\label{Kdiag}
&&\!\!\!\!\!\!{\cal K}\, U^1 = a U^1 + c U^2 \;\; , \;\; {\cal K} \, U^2= c U^1 + b U^2\,, \\
&&\!\!\!\!\!\!{\cal K}\, U^3=\kb U^3 \;,\; {\cal K} U^4 = \kb U^4 \;,\; {\cal K}\, U^5=(\kb+\frac{\|A\|^2}{2}\kd) U^5 \; ,\; {\cal K} \, U^6 = (\kb+\frac{\|A\|^2}{2}\kd) U^6  \,,
\eea
 with
\bea
a=\ka + \kb + \tA^2(\kc + \kd) + (\tA^2)^2 \ke \; , \;\;
b=\kb + 2 \ka \;\; , \;\; c=\sqrt{2}(\ka + \frac{1}{2} \|A\|^2 \kc) \,.
\eea
We thus find that the $6\times 6 $ matrix associated with ${\cal K}$ is  decomposed into a $2\times 2$ matrix and a diagonal $4\times 4$ matrix. Although it is immediate to diagonalize the $2\times 2$ matrix corresponding to the subspace spanned by $(U^1, U^2)$, it is not very useful as we now need to consider the seventh dimension associated  with $V_*$.

\subsection{Mixing with the scalar field}
Interestingly, $V_*$ mixes only with the projection of $K_{ij}$ on the subspace $(U^1, U^2)$, since the mixing coefficient $\B^{ij}$ is of the form 
\bea
{\cal B} = (\beta_1 + \tA^2 \beta_2) U^1 + \sqrt{2} \beta_1 U^2 \,.
\eea
As a consequence, if we decompose $K_{ij}$ according to 
\bea
K_{ij} = K_I\,  U^I_{ij} \;,
\eea
 the kinetic term (\ref{kineticnewbasis}) can be written as
 \bea
{\cal L}_{kin} &=& {\cal A} V_*^2 + V_* \left[ (\beta_1 + \tA^2 \beta_2) K_1 + \sqrt{2} \beta_1 K_2 \right] + aK_1^2 + b K_2^2 + 2c K_1 K_2 
\\
&&+ \kb(K_3^2+K_4^2) +  (\kb+\frac{\|A\|^2}{2}\kd)(K_5^2 + K_6^2) \,.
\eea
We thus find that the kinetic terms along the four directions $U^I$ with $I \in \{ 3,4,5,6\}$, corresponding to the second line above, are trivial. The nontrivial part is embodied by the $3\times 3$ matrix 
\bea\label{degeneratematrix}
\left(
\begin{array}{ccc}
{\cal A} & \frac{1}{2}(\beta_1 + \tA^2 \beta_2) & \frac{1}{\sqrt{2}} \beta_1 \\
\frac{1}{2}(\beta_1 + \tA^2 \beta_2) & a & c \\
\frac{1}{\sqrt{2}} \beta_1 & c & b 
\end{array}
\right),
\eea
which mixes $V_*$ with the metric velocities along $\h_{ij}$ and $A_i A_j$.

\subsection{Degeneracy}
As discussed in detail in our previous paper \cite{Langlois:2015cwa}, one encounters a degenerate theory when  the kinetic part of the action corresponds to a degenerate quadratic form. In general, this degeneracy could arise from the metric kinetic terms, i.e. from ${\cal K}$, if $\kb=0$,  $\kb=-\|A\|^2\kd/2$ or $ab-c^2=0$. However, as we are mainly  interested in theories which conserve two tensor modes, we focus our attention on theories where the degeneracy arises from the mixing with the scalar degree of freedom and we assume that ${\cal K}$ istself is nondegenerate. 

In this case,  ${\cal K}^{ij,kl}$ is invertible and  the degeneracy condition reads~\cite{Langlois:2015cwa}
\bea\label{degeneracy}
{\cal A} - {\cal K}_{ij,kl}^{-1} {\cal B}^{ij} \, {\cal B}^{kl} = 0 \,. 
\eea
It is easy to check that 
this condition is equivalent to the requirement  that the determinant of the $3\times 3$ matrix (\ref{degeneratematrix}) vanishes.

\section{Hamiltonian analysis for degenerate theories}
\subsection{Poisson bracket}
We start the canonical analysis with the definition of the momenta associated to the dynamical variables, via the introduction of the Poisson brackets. 
With the (3+1) decomposition of  the action, we see that the only non-trivial Poisson brackets for the gravitational degrees of freedom are
\begin{eqnarray}\label{variablesgrav}
\{\h_{ij},\pi^{kl}\} = \frac{1}{2} (\delta_i^k \delta_j^l + \delta_i^l \delta_j^k) \;\; , \;\;
\{N,\pi_N\} = 1 \;\; , \;\; \{N^i,\pi_j \} = \delta^i_j \, ,
\end{eqnarray}
and those for the scalar field degrees of freedom are
\begin{eqnarray}\label{variablesphi}
\{\An, p_*\}=1 \;\; , \;\; \{A_i, p^j \} = \delta_i^j \;\; , \;\; \{\phi, p_\phi\} = 1 \, .
\end{eqnarray}
Note that we have identified   the momentum conjugate to $\phi$ with $p_\phi=\lambda^0$. Furthermore, we have not introduced momenta for the variables
$\lambda^i$ which are clearly Lagrange multipliers, and thus  are not dynamical variables. Even if the action does not contain any time derivative of $A_i$, $N$ and $N^i$, we 
cannot a priori consider these variables  as Lagrange multipliers, as they appear non linearly in the action. Nonetheless, we expect 
the lapse $N$ and the shift $N^i$ to be eventually Lagrange multipliers which impose symmetries under diffeomorphisms. As we will see, this is exactly what happens.
\subsection{Primary constraints}
As we have just emphasized, there is no time derivatives of the lapse and the shift in the action. This is due to the introduction of the variable $\An$ which automatically absorbs the time derivatives of these variables. Similarly, there is no time derivative of $A_i$ and $\lambda^i$ is a Lagrange multiplier. 
As a consequence, we get the following 10 primary constraints:
\begin{eqnarray}\label{primary}
\pi_N \approx 0 \;\; , \;\; \pi_i \approx 0 \;\; , \;\; p^i \approx 0 \;\; \text{and} \;\; \chi_i \equiv A_i - D_i\phi \approx 0 \,,
\end{eqnarray}
where the symbol $\approx$ denotes  weak equality (i.e. equality valid on the constraint surface). 
The momenta $\pi^{ij}$ and $p_*$, respectively conjugate to $\h_{ij}$ and $\An$, can be expressed in terms of the configurational variables in the usual way, 
\beq
p_*  =  2\sqrt{\h} ({\cal A} V_* + {\cal B}^{ij}K_{ij} + {\cal C}^0 )\,, \qquad 
\pi^{ij} =   \sqrt{\h} ({\cal K}^{ij,kl} K_{kl} + {\cal B}^{ij} V_* + {\cal C}^{ij}) \,.\label{piij}
\eeq
The degeneracy condition (\ref{degeneracy}) implies that $p_*$ and $\pi^{ij}$ are not independent and satisfy a primary constraint which can be written as
\bea
\label{constraintPsi}
\Psi & \equiv &  p_*  - 2  {\cal K}^{-1}_{ij,kl} {\cal B}^{kl}  \pi^{ij}  +2\sqrt{\h}\left( {\cal K}^{-1}_{ij,kl} {\cal B}^{kl} {\cal C}^{ij} -   2 {\cal C}^0\right) \approx 0\,.
\eea
At this point, we can conclude that,  as ${\cal K}$ is invertible, there is no other primary constraint.

\subsection{Total Hamiltonian}
To go further in the analysis, we need to compute and simplify the expression of the total Hamiltonian $H_{\rm tot}$
defined by
\bea\label{defH}
H_{\rm tot} & = &H + \int d^3x \, (\lambda^i\chi_i + \mu^N \pi_N + \mu^i\pi_i + m_i p^i +m\Psi )\;\;\; \text{with} \cr
H &\approx  & \int d^3x \, (p_*\dot \An + \pi^{ij} \dot{\h}_{ij} + p_\phi \dot\phi ) - L_0 \, .
\eea
Here $\mu^N$, $\mu^i$, $m_i$ and $m$ are Lagrange multipliers enforcing the primary constraints, 
$L_0=L+ \lambda^i\chi_i$ where $L$ is the Lagrangian of the theory (we have suppressed the Lagrange multiplier part
$\lambda^i \chi_i $ to rewrite it explicitly in  the first line above). As a consequence,
\bea\label{Hc}
H &=  & \int d^3x (p_*\dot \An + \pi^{ij} \dot{\h}_{ij} - N \sqrt{\h} {\cal L}_0  + N p_\phi \An + N^i p_\phi A_i) \;\;\; \text{with} \cr
{\cal L}_0 & = &  {\cal A} \, V_*^2 + 2{\cal B}^{ij} V_* K_{ij} + {\cal K}^{ijkl} K_{ij} K_{kl} + 2{\cal C}^{ij} K_{ij} + 2 {\cal C}^0 V_* - \U\, .
\eea
To write the Hamiltonian in terms of the phase space variables,  one needs to reexpress the velocities in terms of the momenta. 
To do so, we first note that, due to the degeneracy condition, the kinetic term in ${\cal L}_0$ factorizes according to
\bea\label{L0simp}
{\cal L}_0 = {\cal K}^{ij,kl}(K_{ij} +{\cal K}^{-1}_{kl,mn}{\cal B}^{mn} V_*)(K_{kl} +{\cal K}^{-1}_{kl,pq}{\cal B}^{pq} V_*) +2 {\cal C}^{ij}K_{ij} +2{\cal C}^0 \dot\An - \U\, .
\eea
Inverting the second relation in  (\ref{piij}) allows to express   $K_{ij}$ in terms of the momenta $\pi^{ij}$ and $\dot\An$,
\bea\label{Kinpi}
K_{ij} = {\cal K}^{-1}_{ij,kl}(\frac{1}{\sqrt{\h}}\pi^{kl} - V_* {\cal B}^{kl} -{\cal C}^{kl})\,,
\eea
which can be substituted in  the Lagrangian density ${\cal L}_0$.
Furthermore, (\ref{Kinpi}) allows to simplify the canonical terms  of the Hamiltonian (\ref{Hc}) as follows:
\bea\label{can}
p_*\dot \An + \pi^{ij} \dot{\h}_{ij} & = & \frac{2N}{\sqrt{\h}} {\cal K}^{-1}_{ij,kl}{\pi^{ij}}{\pi^{kl}} - 2N {\cal K}^{-1}_{ij,kl}{\pi^{ij}}{{\cal C}^{kl}} + 2\pi^{ij} D_i N_j \cr
&&+ NV_*\left((p_*-2 {\cal K}^{-1}_{ij,kl} \pi^{ij}{\cal B}^{kl}\right) 
\eea
Putting everything together, we get the following expression for the Hamiltonian:
\bea\label{H0inter}
H &=  &\int d^3x\,  N \sqrt{\h}\left[ {\cal K}^{-1}_{ij,kl} \left(\frac{\pi^{ij}}{\sqrt{\h}} - {\cal C}^{ij}\right)\left(\frac{\pi^{kl}}{\sqrt{\h}} - {\cal C}^{kl}\right)
+  \U 
\right] +
\cr
 &&   + \int d^3x \left[\Xi \, p_* + N p_\phi \An +N^i \left(p_\phi A_i -2\sqrt{\h} D^j \left(\frac{\pi_{ij}}{\sqrt{\h}}\right) \right)+\sqrt{\h} V_* \Psi\right] \,.
\eea
Note that the dependency on $V_*$ disappears due to the primary constraint $\Psi \approx 0$.  Furthermore, we show, after a direct calculation (and ignoring the surface
terms that appear in the integration by parts), that the Hamiltonian takes the expected form
\bea 
H \approx \int d^3x \, (N{\cal H}_0 + N^i {\cal H}_i)\,,
\eea
with 
\bea 
\label{H0}
{\cal H}_0 & = & \sqrt{\h}\left[ {\cal K}^{-1}_{ij,kl} \left(\frac{\pi^{ij}}{\sqrt{\h}} - {\cal C}^{ij}\right)\left(\frac{\pi^{kl}}{\sqrt{\h}} - {\cal C}^{kl}\right)
+  \U +\frac{p_\phi}{\sqrt{\h}} \An - D_i\left(\tA^i\,  \frac{p_*}{\sqrt{\h}}\right)
\right] \, , \\
{\cal H}_i & = & -2\sqrt{\h}\, D^j \left(\frac{\pi_{ij}}{\sqrt{\h}}\right)  + p_* D_i \An + p_\phi D_i\phi \,. \label{Hdiff}
\eea
Finally, the total Hamiltonian, which defines the time evolution,  reads
\bea
H_{\rm tot}  = \int d^3x \, ( N{\cal H}_0 + N^i {\cal H}_i + \lambda^i\chi_i + \mu^N \pi_N + \mu^i\pi_i + m_i p^i +m\Psi ) \,.
\eea
At this point, it is important to recall that one can replace any of the constraints by a new one which is a 
linear combination of 
the original ones, provided  the new set of constraints remains complete (i.e. the linear transformation between the two sets of constraints is invertible). 
In particular, we can use this property to replace the variables $A_i$ by $D_i\phi$
in each term of $H_{\rm tot}$, except $\chi_i$.  Furthermore, we use  the constraint $\Psi \approx 0$ to replace everywhere, except  in $\Psi$, the momentum
$p_*$ by its expression in terms of $\pi^{ij}$. To avoid heavy notations, we keep the same name for the modified constraints. 
 In conclusion, 
the functions ${\cal H}_0$, ${\cal H}_i$ and $\Psi$ now depend only on the gravitational degrees of freedom $\h_{ij}$ and $\pi^{ij}$, on 
$\An$ and $p_*$ as wellas  on $\phi$ and $p_\phi$, while  the dependence on $A_i$ has been eliminated. 

\subsection{Time evolution of the primary constraints and secondary constraints}
We now study the  time evolution of the primary constraints. Let us recall that the time evolution of any function $F$ defined on the phase space is determined  from the total
Hamiltonian $H_{tot}$, according to
\bea\label{time derivative}
\dot F \equiv \{ F, H_{tot} \} \, .
\eea

\subsubsection{Hamiltonian and momentum constraints}
Let us start with the primary constraints $\pi_N \approx 0$ and $\pi_i \approx 0$. One sees  
immediately that 
\bea
\dot \pi_N \approx 0 \Longrightarrow {\cal H}_0 \approx 0 \;\;\;\; \text{and} \;\;\;\;
\dot \pi_i \approx 0 \Longrightarrow {\cal H}_i \approx 0 \,,
\eea
and it is thus natural to expect that  ${\cal H}_0 \approx 0$ and ${\cal H}_i\approx 0$ correspond to the usual Hamiltonian and momentum constraints of the theory and act as  generators of the 
space-time diffeomorphisms. 

It is easy to check that  ${\cal H}_i$ generates the spatial diffeomorphisms. Its expression (\ref{Hdiff})  is the usual one for a system involving gravity and several scalar fields. More precisely, if one considers the action of the smeared 
function
\beq
{\cal H}(\vec{N})\equiv \int d^3x N^k {\cal H}_k\,,
\eeq
on the variables   $A_*$, $\phi$ and $\h_{ij}$, one easily gets
\bea
\{{A_*, \cal H}(\vec{N})\} = {\cal L}_{\vec{N}} A_*  \;\; , \;\;
\{\phi, {\cal H}(\vec{N})\} = {\cal L}_{\vec{N}} \phi \;\; , \;\;
\{\h_{ij}, {\cal H}(\vec{N})\} = {\cal L}_{\vec{N}} \h_{ij}\,, 
\eea
where ${\cal L}_{\vec{N}}$ is the Lie derivative in the direction $\vec{N}\equiv N^i \partial_i$. 

The action of ${\cal H}(\vec{N})$ on the momenta is slightly different because conjugate momenta are  densities of weight one. For example, $p_*$ transforms as  
\bea
\label{Hi_pn}
\{p_*, {\cal H}(\vec{N}) \} = \partial_i(N^i p_*) \,,
\eea
which is consistent with the fact that  $p_*/\sqrt{\h}$ transforms as a scalar.  Since ${\cal H}_0$ is also a scalar density, its transformation is similarly given by
\beq
\{{\cal H}_0, {\cal H}(\vec{N}) \} = \partial_i(N^i {\cal H}_0) \,.
\eeq

The Poisson brackets of the momentum constraints with themselves are given by
\bea
\{{\cal H}(\vec{N}_1), {\cal H}(\vec{N}_2) \} = {\cal H}(\vec{N})\,, \qquad N^i\equiv N_1^kD_kN_2^i-N_2^kD_kN_1^i \, .
\eea
The Poisson bracket of  ${\cal H}_0$  with itself, 
or possibly of a redefined ${\cal H}_0$ combined  with second-class constraints,
is much more complicated to compute explicitly and we will simply assume, since our starting point is a Lagrangian invariant under four-dimensional diffeomorphisms,
  that it is given by  the usual result
\bea
\{{\cal H}_0(N_1) , {\cal H}_0(N_2) \} = {\cal H}(\vec{N})\,, \quad {\rm with}\quad N^i\equiv N_1 D^k N_2 - N_2 D^k N_1 \, ,
\eea
where 
\beq
{\cal H}_0(N) \equiv \int d^3x \, N\,  {\cal H}_0
\eeq
is the smeared version of the Hamiltonian constraint. We have verified in a simple example  that this is indeed the case (see Appendix \ref{Hamiltonianbracket}).

In conclusion, the time evolution of ${\cal H}_0 \approx 0$ and ${\cal H}_i \approx 0$ does not lead to new, i.e. tertiary, constraints.

\subsubsection{Fixing the Lagrange multipliers $\lambda^i$ and $m_i$}
We now study the  time evolution of the constraints $\chi_i \approx 0$ and $p^i \approx 0$.  The essential ingredient  here is the Poisson bracket
\bea
\{\chi_i, p^j \} = \delta_i^j \, ,
\eea
which immediately implies that
\bea
\dot \chi_i = m_i - D_i (NA_* + N^j D_j \phi)  \;\;\; \text{and} \;\;\; \dot p^i = - \lambda^i \, .
\eea
The time invariance of  the constraints $\chi_i \approx 0$ and $p^i\approx 0$ thus fixes the Lagrange multipliers $m_i$ and $\lambda^i$, 
\bea
 m_i = D_i (NA_* + N^j D_j \phi) \;\;\; \text{and} \;\;\; \lambda^i =0 \, ,
 \eea
 and does not lead to secondary constraints.
 
   It is interesting to notice that the Lagrange multiplier $m_i$
 can be rewritten, according to (\ref{def_An}),  as $m_i=D_i A_0=D_i \dot \phi=\dot A_i$. When we replace this value in the action via the Hamiltonian (\ref{defH}), we obtain a new ``canonical term"
 \bea
 m_i \, p^i = p^i \dot A_i\,,
 \eea
which indicates that $p^i$ and $A_i$ are canonically conjugate variables. This is indeed how the $p^i$ were defined initially, which confirms that  the value  that we get for the Lagrange multiplier
$m_i$ is fully consistent. 

\subsubsection{Secondary constraint from the time evolution of $\Psi$}
It remains to consider the time evolution of the last primary constraint $\Psi \approx 0$.  
Because $\Psi$ commutes with the other primary constraints $\chi_i$, $p^i$, $\pi_i$ and $\pi_N$, its evolution is simply given by 
\bea
\dot \Psi = \{ \Psi , {\cal H}_0(N)  \} + \partial_i(N^i \Psi) \, \approx \, \{  \Psi , {\cal H}_0(N)  \}\,,
\eea
where we have used the property that $\Psi$ is a scalar density and thus transforms like (\ref{Hi_pn}) under the action of ${\cal H}_i$.
We thus  obtain the secondary constraint
\bea
\Omega \equiv \{{\cal H}_0, \Psi \} \, \approx \, 0 \;  \; \; \text{with} \;\;  \Omega =  p_\phi + \Omega_{\rm rest}.
\eea
The explicit form of $\Omega$ is rather involved  in general but we do need its explicit form for our purpose.  What matters is that it depends linearly
on the variable $p_\phi$ as shown above,   as $\Omega_{\rm rest}$ does not contain $p_\phi$. This means that the constraint $\Omega$ can be viewed as an equation that determines the momentum $p_\phi$ in terms of the other variables\footnote{It is not surprising that $p_\phi$ is a redundant variable, since the time derivative of $\phi$ is already contained in the variable $\An$.   The constraint $\Omega$ is the analog,  from the Lagrangian point of view, of the definition of  the momentum $p_\phi$, i.e.
$p_\phi = \partial {\cal L}/{\partial \dot \phi} = {N}^{-1} {\partial {\cal L}}/{\partial \An}$.
 This relation may be rather complicated, in particular for ${\cal A} \neq 0$ where ${\cal L}$ depends  on  $\dot\An$ quadratically, and on $\partial_i\dot \An$ too.
The explicit expression of $p_\phi$ in terms of phase space variables (or in terms of velocities) can thus be quite involved.}.

In order to complete the Dirac analysis, one must then compute the time evolution of $\Omega$, which can be written in the form
\beq
\dot\Omega=\{\Omega, H_{\rm tot}\}=\int d^3 y \, m \, \{\Omega, \Psi\}+\{\Omega,   H_{\rm tot}-\int d^3 y\, m\,  \Psi\}\,,
\eeq
where the second term in the last expression does not depend on $m$.
In the generic case where  $\Delta \equiv \{\Psi,\Omega\}\neq 0$, one thus finds that  the  Lagrange multiplier $m$ is fixed by the time evolution of $\Omega$, which does not generate any new constraint.

In the following, we will not consider the special situations where $\Delta\approx 0$, in which case one expects   a  tertiary constraint or a new symmetry of the theory.  All this would amount to is a further reduction of the physical phase space. This means that  the number of physical degrees of freedom that we are going to compute below,  in the generic case,  can be seen as an upper bound.

\subsection{Number of physical degrees of freedom}
Let us summarize our results. We started with a 30-dimensional phase space, spanned by ten pairs
 of conjugate variables describing the metric, given in  (\ref{variablesgrav}), and five pairs of conjugate variables describing the scalar  field, given in (\ref{variablesphi}).  By performing a  Dirac analysis, we have identified  11 primary constraints (the 10 constraints in (\ref{primary}) and $\Psi \approx 0$, due to the degeneracy) and 5 secondary constraints (${\cal H}_0 \approx 0$,
${\cal H}_i \simeq 0$ and $\Omega \approx 0$). 

As in general relativity, the spacetime diffeomorphism invariance of the initial Lagrangian must translate into the presence of first-class constraints associated with  time and space diffeormorphisms.  
We have showed that the ${\cal H}_i$ indeed generate spatial diffeomorphisms and argued that ${\cal H}_0$,   possibly combined with second class constraints, should correspond to the Hamiltonian constraint that generates time reparametrisation. Furthermore, as none of the constraints depend on the lapse $N$ and on the shift $N^i$, $\pi_N \approx 0$ and $\pi_i \approx 0$ are necessarily first-class constraints as well. We thus have 8 first-class constraints.

The remaining 8 constraints $\Phi_A=(p^i, \chi_i , \Psi , \Omega )$ form a family of second-class constraints, as we now show. We first recall  that we have used of $\chi_i \approx 0$ to replace the variables $A_i$ by $\partial_i \phi$ in all the constraints, except  of course $\chi_i$. With this in mind,  it is immediate to see that the non-vanishing
components of the Dirac matrix $M_{AB}(x,y)\equiv\{\Phi_A(x),\Phi_B(y)\}$  are given by
\bea
\{ \chi_i(x) , p^j(y) \} = \delta_i^j \, \delta(x-y)\; , \; \{\Psi(x),\Omega(y) \} = \Delta\,  \delta(x-y)  \,,\, \{\chi_i(x),\Omega(y) \}= - \partial_{x^i}\delta(x-y)\,,
\eea
where we have made manifest the spatial dependence,  due to the presence of the derivative of $\delta(x-y)$. 
Since $\Delta \neq 0$, the Dirac matrix is clearly invertible which means that $\Phi_A$  are second-class constraints. These constraints allow  to eliminate the variables $p^i$ and $A_i$ and to reexpress  $p_*$ and $p_\phi$  in terms of $\h_{ij}$, $\pi^{ij}$, $\phi$ and $\An$ only. All other variables are redundant and can be eliminated by solving secondary constraints.

To conclude, let us compute the number of physical degrees of freedom. The dimension of the physical phase space is given by 30 - 2$\times$(number of first class constraints) 
- (number of second class constraints)= $30-2 \times 8 - 8= 6$, which gives three degrees of freedom. As expected, this corresponds to two tensor modes and only one scalar degree of freedom. This confirms that the extra scalar degree of freedom associated with the Ostrogradski instability is not present in degenerate scalar-tensor theories.

\section{Nondegenerate theories}
\label{nondegenerate}
For completeness, let us  turn to the case of nondegenerate theories. We reproduce the procedure followed in the previous section, starting with the action (\ref{3+1decomp}).

As before, our pairs of conjugate variables are defined by (\ref{variablesgrav}) and (\ref{variablesphi}).  The nondegeneracy of the Lagrangian, assumed in this section,  implies that the relations
between the momenta ($p_*$, $\pi^{ij}$) and the velocities, namely
\bea
\left(
\begin{array}{c}
\frac{p_*}{2\sqrt{\h}}-\C^0 \\
\frac{\pi^{ij}}{\sqrt{\h}}-\C^{ij}
\end{array}
\right)
 =  
\left(
\begin{array}{cc}
{\cal A} & {\cal B}^{kl} \\
{\cal B}^{ij} & {\cal K}^{ij,kl}
\end{array}
\right)
\left(
\begin{array}{c}
V_*  \\
K_{kl}
\end{array}
\right) \,,
\eea
can be inverted. This can be done explicitly by introducing the inverse of the kinetic matrix, 
\bea
\left(
\begin{array}{cc}
{\cal A} & {\cal B}^{kl} \\
{\cal B}^{ij} & {\cal K}^{ij,kl}
\end{array}
\right)^{-1}
\equiv 
\left(
\begin{array}{cc}
\hat{\cal A} & \hat{\cal B}_{kl} \\
\hat{\cal B}_{ij} & \hat{\cal K}_{ij,kl}
\end{array}
\right)
\; \text{with} \;
\left\{
\begin{array}{l}
\hat{\cal A}= ({\cal A} - {\cal K}^{-1}_{ij,kl} {\cal B}^{ij} {\cal B}^{kl})^{-1}\\
\hat{\cal B}_{ij}=-\hat{\cal A} \, {\cal K}^{-1}_{ij,kl} {\cal B}^{kl}\\
\hat{\cal K}_{ij,kl}=   {\cal K}^{-1}_{ij,kl}  +\hat{\cal A}^{-1} \,  \hat{\cal B}_{ij} \hat{\cal B}_{kl}
\end{array}
\right.
.
\eea
Here there is no primary constraint between $p_*$ and $\pi_{ij}$ and the set of primary constraints reduces to (\ref{primary}). The total Hamiltonian of the theory is thus given by 
\bea
H_{\rm tot} & = & \int d^3x (p\dot \An + \pi^{ij} \dot{\h}_{ij} - N \sqrt{\vert \h \vert} {\cal L}_0  + N p_\phi \An + N^i p_\phi A_i)  \cr
&  + &    \int d^3x (\lambda^i \chi_i + \mu^N \pi_N + \mu^i \pi_i + m_i p^i) \,.
\eea
A straightforward calculation easily leads  to the following expression for the total Hamiltonian:
\bea
H_{\rm tot} & = & \int d^3x \, ( N{\cal H}_0 + N^i {\cal H}_i + \lambda^i\chi_i + \mu^N \pi_N + \mu^i\pi_i + m_i p^i) \,,
\eea
where the Hamiltonian constraint ${\cal H}_0$ and the momentum constraints ${\cal H}_i$ are given by:
\bea
{\cal H}_0 & = & \frac{1}{\sqrt{\h}}(\frac{1}{4} \hat{\cal A} \, p_*^2 +  \hat{\cal B}_{ij} p_* \pi^{ij}  + \hat{\cal K}_{ij,kl} \pi^{ij} \pi^{kl}) + \cr 
&&+   \An p_\phi-\left(\hat{\cal A} \C^0+\hat{\cal B}_{ij} \C^{ij}\right) p_*
- 2\left(\C^0 \hat{\cal B}_{kl} +\C^{ij}\hat{\cal K}_{ij,kl}\right)\pi^{kl} -\sqrt{\h}D_i\left(\frac{p_*}{\sqrt{\h}}D^i\phi\right)  \cr 
&&
+ \sqrt{\h} \left[\U+2 C^0 \hat{\cal B}_{ij} \C^{ij}+\hat{\cal K}_{ij,kl} \C^{ij}\C^{kl}\right]
\label{H0_nd} \, ,\\
{\cal H}_i & = & -2\sqrt{\h}D^j \left(\frac{1}{\sqrt{\h}}\pi_{ij}\right) + p_*D_i \An + p_\phi D_i \phi\, .
\eea

The analysis of the constraints is easier than in the degenerate case. Stability under time evolution of the constraints $\pi_N \approx 0$
and $\pi_i \approx 0$ leads to the secondary constraints which are the Hamiltonian and vectorial constraints. It is immediate to see that
${\cal H}_i$ is first-class and generates space diffeomorphims. As for ${\cal H}_0$, we show in Appendix \ref{Hamiltonianbracket} that 
it satisfies the expected Poisson algebra in a simple example. We expect this to be true in general and we thus assume that ${\cal H}_0$, combined with second-class constraints,  is first-class.
 Stability under time evolution of $\chi_i \approx 0$ and $p_i \approx 0$ leads to fixing the 
Lagrange multipliers $m_i$ and $\lambda^i$.

In summary, the theory admits $30$ non-physical degrees of freedom ($\h_{ij}$, $N$, $N^i$, $\An$, $\phi$, $A_i$ and their momenta) 
for $14$ constraints. The constraints are divided  into $8$ first-class constraints (${\cal H}_0$, ${\cal H}_i$, $\pi_N$ and $\pi_i$) and $6$ second-class
constraints ($p^i$ and $\chi_i$). Thus, the theory admits $4$ degrees of freedom ($8$ degrees of freedom in the phase space)  
which correspond to 2 tensorial degrees of freedom, 1 scalar and 1 ghost. 

Exactly as in the degenerate case, we can solve explicitly the $6$ second-class constraints by replacing everywhere in the theory $A_i$ by $\phi_i$ and eliminating
$p^i$. Furthermore, we can  consider the lapse $N$ and the shift $N^i$ as Lagrange multipliers  and thus 
eliminate the constraints $\pi_N \simeq 0$ and $\pi_i \simeq 0$.  Finally, we end up with a theory that can be formulated in terms of $16$ degrees of freedom 
in phase space $(\phi,\An,\h_{ij}$ and their momenta) which satisfy $4$ first-class constraints ${\cal H}_0\approx 0$ and ${\cal H}_i\approx 0$.

\section{On the unitary gauge}
In this section, we focus on  the unitary gauge, which was used in the early Hamiltonian analyses of the theories beyond Horndeski~\cite{Gleyzes:2014dya,Lin:2014jga,Gleyzes:2014qga}, because of its simplicity.  We will show that, in general, the unitary gauge is a good gauge,  which
breaks time reparametrization. However, there exist particular situations where the unitary gauge is not allowed because it leads to a singular Hamiltonian  in the  phase space region where the unitary gauge is imposed.
These cases correspond to theories that are nondegenerate but look degenerate in the unitary gauge, such as those discussed in the Appendix of \cite{Langlois:2015cwa}.

The unitary gauge consists in choosing the scalar field as the clock. More concretely, we impose a new primary constraint  given by
\bea\label{gauge fixing}
{\cal F} \; \equiv \; \phi - t \; \approx \; 0 \, .
\eea
We could have imposed $\phi$ to be an arbitrary monotonous function $f(t)$, but for simplicity we make the choice $f(t)=t$. 
Gauge fixing means that we consider a new theory with the total Hamiltonian 
\bea\label{gaugefixed}
H_{\rm tot}^{\rm gauge} \; \equiv \; H_{\rm tot} + \int d^3 x \, \xi\,  {\cal F} \,,
\eea
where the Lagrange multiplier $\xi$  enforces the gauge fixing (\ref{gauge fixing}). To see whether this is indeed an appropriate gauge fixing,  we  repeat  the analysis of the constraints and verify that the symmetry under time reparametrization is indeed broken.

\subsection{Unitary gauge in nondegenerate theories}
Let us first consider nondegenerate theories with the Hamiltonian
\bea
H_{\rm tot} & = & \int d^3x \, ( N{\cal H}_0 + N^i {\cal H}_i + \lambda^i\chi_i + \mu^N \pi_N + \mu_i\pi^i + m_i p^i) \,.
\eea
In the corresponding gauge fixed Hamiltonian (\ref{gaugefixed}), one can replace ${\cal H}_0$ and ${\cal H}_i$ by their respective expressions with $\partial_i\phi =0$.   Similarly, $\chi_i$ can be replaced by $\chi_i = A_i$. One thus finds that the expression of the total Hamiltonian simplifies drastically. 

Stability under time evolution of $\chi_i$ and $p^i$ fixes both Lagrange multipliers $\mu_i$ and $\lambda^i$,  as  in the  arbitrary gauge case. 
To see the effect of the gauge fixing, 
it is sufficient to study the time evolution of ${\cal F}$. 
As ${\cal F}$ depends explicitly on time, its time derivative is now given by
\bea
\dot {\cal F} \equiv  \frac{\partial {\cal F}}{\partial t} +  \{  {\cal F} , H_{\rm tot}^{\rm gauge} \} \approx N \An -1  \,,
\eea
which leads to the new secondary constraint
\bea
{\cal G} \equiv N\An - 1 \approx 0 \, .
\eea
Requiring the time invariance of  ${\cal G}$  fixes the Lagrange multiplier $\mu^N$.
An analysis  of the time evolution of the other constraints shows that they do not produce  additional constraints. 

Let us now examine the nature (first or second class) of the constraints. 
As expected, ${\cal H}_i \approx 0$ and $\pi_i \approx 0$
remain first-class because the invariance under space diffeomorphims is not broken. 
By contrast, ${\cal H}_0 \approx 0$ and $\pi_N \approx 0$ are no longer first-class and together
with ${\cal F} \approx 0$ and ${\cal G} \approx 0$, $\chi_i \approx 0$ and $p^i\approx 0$, they form a set of second-class constraints. 
Note that $p^i$ and $\chi_i$ commute with the four others and therefore can be treated separately (in fact, they can be solved explicitly and thus be ignored in the following).
The Dirac matrix $M_{AB}= \{\Phi_A,\Phi_B\}$ associated with the remaining four constraints (with $\Phi_1\equiv {\cal H}_0$, $\Phi_2 \equiv \pi_N$, $\Phi_3\equiv {\cal F}$, 
$\Phi_4\equiv {\cal G}$) is given by
\bea
\left(
\begin{array}{cccc}
 0&0 & -1/N & \{\Phi_1, \Phi_4\} \\
 0 & 0 & 0 & -1/N \\
 1/N & 0 & 0 & 0\\
-\{\Phi_1, \Phi_4\} & 1/N & 0 & 0
\end{array}
\right)
\eea
because 
\bea
\{{\cal H}_0 , {\cal F}\} \approx -\frac{1}{N} \;\;\; , \;\;\;
\{{\cal H}_0 , {\cal G} \} = N \{{\cal H}_0,\An \}  \;\;\; ,\;\;\;
\{{\cal G}, \pi_N \} \approx - \frac{1}{N} \,.
\eea
For any finite value of the Poisson bracket which simplifies in the unitary gauge to
\bea
\{{\cal H}_0,\An\} \, \approx \, \frac{1}{\sqrt{\h}} (\frac{1}{2N} \hat{\cal A} p_* + \hat{\cal B}_{ij} \pi^{ij})
\eea
the Dirac matrix $M_{\alpha\beta}$ is invertible. 

In conclusion, the above analysis shows that  the unitary gauge ${\cal F} \approx 0$ is, in general, a valid gauge.
However, it supposes that  the Hamiltonian $H_{\rm tot}$ itself is  well defined in the unitary gauge. When the coefficient ${\cal A} - {\cal K}^{-1}_{ij,kl} {\cal B}^{ij} {\cal B}^{kl}$ vanishes in the unitary gauge, even if the theory is nondegenerate,  then the coefficient $\hat\A$ is infinite and the Hamiltonian becomes singular in the unitary gauge. It means that the unitary gauge is problematic for theories  that are nondegenerate but look degenerate when restricted to $\partial_i\phi=0$.

\subsection{Unitary gauge in degenerate theories}
For degenerate theories, the analysis of the unitary gauge is similar but a bit  subtler due to the presence of the extra primary constraint
$\Psi \approx 0$ (which generates the secondary constraint $\Omega \approx 0$). 
For that reason, we will give more details than in the non-degenerate case.

The  Hamiltonian that appears in (\ref{gaugefixed}) is
\bea
H_{\rm tot} & = & \int d^3x \, ( N{\cal H}_0 + N^i {\cal H}_i + \lambda^i\chi_i + \mu^N \pi_N + \mu_i\pi^i + m_i p^i + m\Psi) \,.
\eea
The constraints ${\cal H}_0$, 
${\cal H}_i$  and $\Psi$, given in (\ref{H0}), (\ref{Hdiff}) and  (\ref{constraintPsi}) respectively, simplify into
\bea 
{\cal H}_0 & = & \frac{1}{\kb \sqrt{\h}} \left( \pi^{ij} \pi_{ij} - \frac{\ka}{\kb + 3\ka} \pi^2\right) +\An \left( \frac{2 f_{,\phi}}{\kb+3\ka}  \pi + p_\phi \right) + \sqrt{\h} \left( \frac{3 {\omega}^2}{\kb + 3\ka} + {\cal U} \right)  \\
{\cal H}_i & = & -2 D^j {\pi_{ij}}   + p_* D_i \An \\
\Psi & = &  p_*  - \frac{2{\ba}}{\kb + 3\ka}  \left( \pi +3 \sqrt{\h}f_{,\phi} \An \right)\,,
\eea
where $\pi=\pi^{ij} \gamma_{ij}$ is the trace of the momentum  and we have used 
\bea
{\cal K}^{ij,kl} & = & \kb \gamma^{i(k} \gamma^{l)j} + \ka \gamma^{ij} \gamma^{kl} \,\,\, \text{and} \,\,\, 
{\cal K}^{-1}_{ij,kl}  =  \frac{1}{\kb} \gamma_{i(k} \gamma_{l)j} -\frac{\ka}{\kb(\kb + 3\ka)} \gamma_{ij} \gamma_{kl} \\
{\cal U} & = & - R - 4D_i(f_{,X} \An D^i \An) - (\ad \An^2 - 2\ab)(D_i \An)(D^i \An) \\
{\cal B}^{ij} & = & \ba \gamma^{ij}  \;\;\; \text{and} \;\;\; {\cal C}^{ij} =  -f_{,\phi} \An \gamma^{ij}  \, .
\eea
Let us note that the unitary gauge can be used  only if ${\cal K}^{ij,kl}$ remains invertible in this gauge, which means that
$\kb$ and $\kb+3\ka$ must be non-zero. When  ${\cal K}^{ij,kl}$ becomes degenerate in the unitary gauge, then the Hamiltonian 
is ill-defined and thus the gauge is not safe. This happens when 
\bea
\ab X - f = 0 \;\;\; \text{or} \;\;\; (\ab + 3�\aa)X + 2  f = 0\,,
\eea
where we have used that  $\An^2=-X$ in the unitary gauge.  

The analysis of secondary constraints is similar to the non-degenerate case with the difference that time evolution of $\Psi$ leads to 
the secondary constraint $\Omega \approx 0$ as expected. We still have the vectorial constraints ${\cal H}_i \approx 0$ and $\pi^i \approx 0$
which form a set of first-class constraints. It remains to study the following 6 constraints, which we denote $\Phi_A \approx 0$:
\bea
\Phi_1={\cal H}_0 \;, \;\;
\Phi_2={\pi}_N  \;, \;\;
\Phi_3={\cal F} \;, \;\;
\Phi_4={\cal G}  \;, \;\;
\Phi_5={\Psi}  ,\;\;
\Phi_6=\Omega  \, .
\eea
They must be second-class to ensure that  the unitary gauge is applicable.  We thus need 
to compute the determinant of the full Dirac matrix $\{\Phi_A,\Phi_B\}$,  which is weakly equal to 
\bea
\left(
\begin{array}{cccccc}
0 & 0 & -1/N & 0 & 0 & \{\Phi_1, \Phi_6\} \\
0 & 0 & 0 & -1/N & 0 & 0 \\
 1/N & 0 & 0 & 0 & 0 & 1 \\
 0 & 1/N & 0 & 0 & N & 0 \\
 0 & 0 & 0 & -N & 0 & -\Delta \\
 \{\Phi_6, \Phi_1\} & 0 & -1 & 0 & \Delta & 0 
\end{array}\,.
\right)
\eea 
An immediate calculation shows that its determinant is $\Delta^2/N^4$ which is nonzero when $\Delta \neq 0$
as we assumed at the begining. This confirms that the unitary gauge is an appropriate gauge, provided the Hamiltonian is well defined. 

\section{Conclusions}
In this work, we have presented a Hamiltonian formulation of higher order theories of the form (\ref{generalHorn}), both in the degenerate and nondegenerate cases. The degenerate case is especially important as it includes the quadratic Horndeski Lagrangian $L_4^{\rm H}$, as well as its extension beyond Horndeski $L_4^{\rm bH}$. 

By using  the variables  introduced in our previous work, we have been able to compute the total Hamiltonian for degenerate and nondegenerate theories. In both cases, our Hamiltonian is linear in the lapse $N$ and the shift $N^i$ and thus reproduces the familiar structure of the GR Hamiltonian, enabling us to identify the Hamiltonian and momentum first-class constraints associated with the invariance under spacetime diffeomorphisms. 
The only caveat in our derivation is that we did not compute explicitly the Poisson brackets of the Hamiltonian constraint with itself in order to check the full recovery of the familiar algebra. Or, more precisely, we checked it only for a simple (nondegenerate) theory, where the brute force calculation is already quite involved.  
However,  it is natural to  believe that this result should be true in general. 

Our analysis confirms the conjecture of our previous paper that degenerate theories of the form (\ref{generalHorn})
should contain only three dynamical degrees of freedom whereas the dynamics of their nondegenerate counterparts should include an extra scalar degree of freedom, which is expected to behave as an Ostrogradski ghost. 
To our knowledge, this is the first derivation of the Hamiltonian formulation for the quadratic Horndeski Lagrangian $L_4^{\rm H}$, confirming the absence of an Ostrogradski ghost. In the special case of  the Lagrangian $L_4^{\rm bH}$, our Hamiltonian formulation appears rather simpler than the one  presented in \cite{Deffayet:2015qwa}, based on a completely different choice of canonical variables. Furthermore, our analysis also applies to the new degenerate theories identified in our previous work. 

In the future, it would be interesting to extend the present results to a larger class of theories, in particular theories which are cubic in second derivatives of the scalar field, such as the quintic Horndeski and beyond Horndeski Lagrangians. However, the difficulty to invert explicitly the relation between the momenta and velocities might be an obstacle in practice. It would also be instructive to clarify  the relation between the number of degrees of freedom and  the order of  the equations of motion. As a first step,  it would  be easier  to study this question  in the context of  particle mechanics~\cite{mnsyl}.

\acknowledgements
We would like to thank Marco Crisostomi, Gilles Esposito-Farese and Daniele Steer for interesting discussions.

\appendix
\section{Elimination of the time derivatives $\dot A_i$ in the Lagrangian}
\label{Appendix_A}
As mentioned in the main text, one can use the property $\nabla_\mu A_\nu = \nabla_\nu A_\mu$, which directly follows from the relation $A_\mu=\nabla_\mu\phi$ to replace all the terms $\nabla_0 A_i$  by $\nabla_i A_0$ in the action. 
Indeed, whenever  one encounters an expression of the form  $B^i \nabla_0 A_i$ in the Lagrangian, where $B^i$ is an arbitrary  combination of the variables, one can always write
\begin{eqnarray}\label{symmetry}
\int  d^4x \, B^i \nabla_0 A_i & = & \int  d^4x \,  B^i  \left[ \nabla_0 (A_i - \phi_i)  + \nabla_0 \nabla_i \phi \right] \\
& = & \int  d^4x \,  B^i  \left[ \nabla_0 (A_i -  \phi_i)  + \nabla_i (\phi_0 - A_0) + \nabla_i A_0 \right] \\
& = &  \int  d^4x \,\left[  -(\nabla_0 B^i)   (A_i -  \phi_i)  - (\nabla_i B^i) (\phi_0 - A_0) + B^i \nabla_i A_0 \right]\,,
\end{eqnarray}
where the last line is obtained via an integration by parts, leaving aside the boundary terms. Finally, after a redefinition of the variables $\lambda^\mu$,  one can check that the Lagrangian is unaffected by this change. We thus conclude that all the time derivatives of the spatial components $A_i$ can be eliminated in the Lagrangian.

\section{ADM decomposition of $\nabla_\mu A_\nu$}
\label{Christoffel appendix}
Here, we  compute the components of the covariant derivative 
\bea
\nabla_\mu A_\nu \equiv \partial_\mu A_\nu - \Gamma_{\mu \nu}^\rho \, A_\rho
\eea
using the expressions of the  Christoffel symbols $\Gamma_{\mu\nu}^\rho$ in term of ADM quantities. They are given by
\begin{eqnarray}
\Gamma_{00}^0&=&\frac1N\left(N^iN^j K_{ij}+\dot N+N^i D_iN\right) \; ,
\cr
\Gamma_{00}^k&=&N N^i\left(2\h^{jk} -\frac{N^jN^k}{N^2}\right)K_{ij}+\dot N^k-\frac{N^k}{N}\dot N+N^i D_iN^k+N\left(\h^{kl}-\frac{N^kN^l}{N^2}\right)D_lN  \; ,
\cr
\Gamma_{0i}^0&=&\frac1N\left(N^kK_{ki}+D_iN\right)  \; ,
\cr
\Gamma_{0i}^j&=& N\left(\h^{jk}-\frac{N^j N^k}{N^2}\right)K_{ik}+D_iN^j-\frac{N^j}{N}D_iN  \; ,
\cr
\Gamma_{ij}^0&=& \frac1N K_{ij}  \; ,
\cr
\Gamma_{ij}^k&=& -\frac{N^k}{N} K_{ij}+\hat\Gamma_{ij}^k\,.
\end{eqnarray}
In the last equation,  $\hat\Gamma_{ij}^k$ denote the three-dimensional Christoffel symbols associated to the spatial metric $\h_{ij}$. 
From these expressions,  one can easily obtain the different components of the covariant derivative of $A_\mu$
\begin{eqnarray}
A_{00}&\equiv& \nabla_0A_0=\dot A_0-\Gamma_{00}^0A_0-\Gamma_{00}^k A_k
=N\dot\An -\left(\An  N^i N^j+2N A^{(i} N^{j)}\right)K_{ij} 
\cr
&&\qquad +N N^kD_k\An +N^iN^jD_iA_j-NA^kD_kN +N^k(\dot A_k-D_k A_0)  \; ,\label{nabla00}
\\
A_{i0}&\equiv& \nabla_iA_0=D_i A_0-\Gamma_{0i}^0A_0-\Gamma_{0i}^k A_k=
-(\An  N^j+NA^j)K_{ij}
\cr
&&\qquad +ND_i\An +N^kD_iA_k   \; ,
\\
A_{0i}&\equiv& \nabla_0A_i=\dot A_i-\Gamma_{0i}^0A_0-\Gamma_{0i}^k A_k= (\dot A_i-D_i A_0)
-(\An  N^j+NA^j)K_{ij}
\cr
&&\qquad +ND_i\An +N^kD_iA_k  \; , \label{nabla0i}
\\
A_{ij}&\equiv& \nabla_i A_j=\partial_i A_j-\Gamma_{ij}^0A_0-\Gamma_{ij}^k A_k
= D_i A_j-\An  K_{ij}\,.
\end{eqnarray}
These expressions can also be directly obtained by projecting the 3+1 covariant decomposition of  $\nabla_\mu A_\nu$, given in \cite{Langlois:2015cwa}, onto a basis associated with the coordinates $t$ and $x^i$.

\section{Poisson bracket $\{{\cal H}_0, {\cal H}_0\}$}
\label{Hamiltonianbracket}
This  goal of this Appendix is to verify that
\bea
\{{\cal H}_0(N_1) , {\cal H}_0(N_2)\} = (N_1 D_i N_2 - N_2 D_i N_1) {\cal H} ^i\,,
\eea 
for the  special case
\beq
S[g_{\mu\nu},\phi] \equiv \int d^4x \, \sqrt{\vert g \vert} \left(  {\cal R} +  \alpha (\nabla_\mu \nabla_\nu\phi) \, (\nabla^\mu \nabla^\nu\phi) \right) \,,
\eeq
where $\alpha$ is assumed to be constant. Note that this theory, which  is of the form (\ref{family}) with $f=1$,  $\alpha_1= \alpha_2= \alpha_3 = \alpha_4 = 0$ and $\alpha_5 = \alpha$, is nonegenerate.

According to (\ref{H0_nd}), the smeared  constraint ${\cal H}_0(N)$ is explicitly given by 
\bea
{\cal H}_0(N) & = & \int d^3x \, N \left[ \frac{1}{\sqrt{\gamma}} (\frac{1}{4\alpha} p_*^2 + {\cal K}^{-1}_{ij,kl} \pi^{ij} \pi^{kl}) - 2 {\cal C}^{ij} {\cal K}^{-1}_{ij,kl} \pi^{kl} 
+\sqrt{\gamma} (\U + {\cal K}^{-1}_{ij,kl} {\cal C}^{ij} {\cal C}^{kl}) + \An p_\phi\right] \cr \nonumber \\
&& +\int d^3x \, (D_i N) p_* \gamma^{ij} A_j \, 
\eea
with
\bea
{\cal K}^{ij,kl} & = & (1+ \alpha \An^2) \gamma^{i(k} \gamma^{l)j} - \gamma^{ij} \gamma^{kl} - \alpha (\tA^i \tA^{(k} \gamma^{l)j} + \tA^j \tA^{(k} \gamma^{l)j}) \, , \\
{\cal C}^{ij} & = & -\alpha \An D^i \tA^j + \alpha(\tA^i D^j \An + \tA^j D^i \An) \, ,\\
{\U} & = & - R - \alpha (D_i A_j) (D^i \tA^j) + 2\alpha (D_i \An)(D^i \An) \, .
\eea
The Poisson bracket we wish to compute is given by 
\bea\label{HH}
\{ {\cal H}_0(N_1) , {\cal H}_0(N_2)\}  & = & \int d^3 x \left[ \frac{\delta {\cal H}_0(N_1)}{\delta \An } \frac{\delta {\cal H}_0(N_2)}{\delta p_*}  +
 \frac{\delta {\cal H}_0(N_1)}{\delta \phi} \frac{\delta {\cal H}_0(N_2)}{\delta p_\phi} +
  \frac{\delta {\cal H}_0(N_1)}{\delta \gamma_{ij} } \frac{\delta {\cal H}_0(N_2)}{\delta \pi^{ij}} 
 \right] \nonumber \\
 & - &   (N_1 \leftrightarrow N_2) \, 
\eea
where $(N_1 \leftrightarrow N_2)$ means that one exchanges the role of $N_1$ and $N_2$ in the first line.
For that purpose, we need to compute derivatives of ${\cal H}_0(N)$ with respect to the various phase space variables.

Derivatives with respect to the momenta are easy to compute:
\bea
\frac{\delta {\cal H}_0(N)}{\delta p_*} & = & N \frac{1}{2\alpha \sqrt{\gamma}} p_* + (D^i N) A_i \\
 \frac{\delta {\cal H}_0(N)}{\delta p_\phi} &= & N \An \label{wrtpphi}\\
  \frac{\delta {\cal H}_0(N)}{\delta \pi^{ij}} & = & {2N}  {\cal K}^{-1}_{ij,kl} (\frac{\pi^{kl}}{\sqrt{\gamma}} -{\cal C}^{kl} ) \label{wrtpi}
\eea
Derivatives with respect to the variables $\An$, $\phi$ and $\gamma_{ij}$ are more involved to compute, and it is useful to derive some intermediate results.
Let us start with the derivatives with respect to $A_i$ and $A_*$ of  the coefficients that appear in  ${\cal H}_0(N)$. 
For 
${\cal K}^{-1}_{ij,kl}$, we get
\bea
\frac{\partial {\cal K}^{ij,kl}}{\partial A_m} & = & -\alpha [\gamma^{im} \tA^{(k} \gamma^{l)j} + \gamma^{jm} \tA^{(k} \gamma^{l)i} 
+ \tA^i \gamma^{m(k} \gamma^{l)j} +  \tA^j \gamma^{m(k} \gamma^{l)i}] \\
\frac{\partial {\cal K}^{-1}_{ij,kl}}{\partial \An} & = & - 2\alpha \An {\cal K}^{-2}_{ij,kl} \\
\frac{\partial {\cal K}^{-1}_{ij,kl}}{\partial A_m} & = & 2\alpha \tA^n \left[ {\cal K}^{-1}_{ij}{}^{mp} {\cal K}^{-1}_{np,kl} +  {\cal K}^{-1}_{kl}{}^{mp} {\cal K}^{-1}_{np,ij} \right]
\eea
while the derivatives of ${\cal C}^{ij}$ are given by
\bea
\frac{\partial {\cal C}^{ij}}{\partial \An} & = & - \alpha D^i \tA^j\,,\qquad \frac{\partial {\cal C}^{ij}}{\partial D_m \An} =  \alpha \left[ \tA^i \gamma^{jm} + \tA^j  \gamma^{im}\right] \\
\frac{\partial {\cal C}^{ij}}{\partial A_m} & = & \alpha \left[ \gamma^{mi} D^i \An + \gamma^{mj}D^i \An \right]\,, \qquad 
\frac{\partial {\cal C}^{ij}}{\partial D_mA_{n}}  =  -\alpha \An \gamma^{i(k} \gamma^{l)j} 
\eea
and  the derivatives of $\U$ by
\beq
\frac{\partial {\U}}{\partial D_mA_{n}}  =  -2\alpha D^{m}\tA^n\,, \qquad
\frac{\partial {\U}}{\partial D_m\An}  =  4\alpha D^{m}\tA^n\,.
\eeq

Only terms which depend on  derivatives of the metric will enter the Poisson bracket (\ref{HH}). Indeed, ${\cal H}_0(N)$ does not depend on the derivatives of the momenta $\pi^{ij}$ and  (\ref{HH}) is antisymmetric in the exchange $N_1 \leftrightarrow N_2$.
Derivatives of the metric appear only in $\U$ through  the 3 dimensional Ricci scalar  $R$ and $(D_i A_j)(D^i \tA^j)$, and also  in ${\cal C}^{ij}$ through the
covariant derivatives of $A_i$. Thus, we only need the following  formulae:
\bea
\frac{\delta }{\delta \gamma_{ij}} \int d^3x N \sqrt{\gamma} R & = & N [\cdots]^{ij} + \sqrt{\gamma} \left[ D^i D^j N - \gamma^{ij} D^m D_m N \right] \\
\frac{\delta}{\delta \gamma_{ij}} \int d^3x N \sqrt{\gamma} (D_k A_l)^2 & = & N [\cdots]^{ij} + \sqrt{\gamma} (D_k N)[ \tA^i (D^j \tA^k) +  \tA^j (D^i \tA^k) - \tA^k D^i\tA^j] \\
\frac{\delta}{\delta \gamma_{ij}}  \int d^3x \, N \Theta_{kl} \, {\cal C}^{kl} & = & N [\cdots]^{ij} + \frac{\alpha}{2} \An (D_k N) [ \tA^k \Theta^{ij} -  \Theta^{jk} \tA^i  -\Theta^{ik} \tA^j]
\eea
where $\Theta_{ij}$ is any tensor independent of derivatives of $\gamma_{ij}$. Terms proportional to $N$ are not relevant for the calculation of
(\ref{HH}) and we do not need their explicit form.

Gathering the above results together, we obtain for ${\delta {\cal H}_0(N)}/{\delta \An} $ the  expression
\bea\label{D1}
\frac{\delta {\cal H}_0(N)}{\delta \An} & = & \frac{\partial {\cal H}_0(N)}{\partial \An} - D_i \left[ \frac{\partial {\cal H}_0(N)}{D_i \partial \An} \right] \cr
& =& N p_\phi  + 2\alpha N \sqrt{\gamma} 
\left[ -{\An}  {\cal K}^{-2}_{ij,kl}(\frac{\pi^{ij} }{\sqrt{\gamma}} -  {\cal C}^{ij}) (\frac{\pi^{kl}}{\sqrt{\gamma}}  - {\cal C}^{kl}) \right.  \cr
& + &  \left. {\cal K}^{-1}_{ij,kl} (D^i\tA^j) (\frac{\pi^{kl}}{\sqrt{\gamma}} - {\cal C}^{kl}) + 2 D_j [A_i {\cal K}^{-1ij}_{kl} (\frac{\pi^{kl}}{\sqrt{\gamma}} - {\cal C}^{kl})]
-2 D_i D^i \An \right] \cr
& + & + 4 \alpha \sqrt{\gamma} (D_j N) \left[ A_i {\cal K}^{-1ij}_{kl} (\frac{\pi^{kl}}{\sqrt{\gamma}} - {\cal C}^{kl}) - D^j \An \right] \,.
\eea
For  the two other  derivatives, their component proportional to the lapse $N$ does not contribute to the Poisson bracket (\ref{HH}) because
(\ref{wrtpphi}) and (\ref{wrtpi}) are proportional to $N$. Thus, we concentrate only on the terms proportional to derivatives of the lapse and we obtain
\bea\label{D2}
\frac{\delta {\cal H}_0(N)}{\delta \phi} & = & N [\cdots] -(D_i N)  \sqrt{\gamma} D^i (\frac{p_*}{\sqrt{\gamma}}) -  (D_i D^i N) p_* \cr
& - & 4\alpha \sqrt{\gamma} (D_m N) \left[ {\cal K}^{-1}_{ij}{}^{ms} {\cal K}^{-1}_{rs,kl} \tA^r (\frac{\pi^{ij}}{\sqrt{\gamma}} - {\cal C}^{ij}) (\frac{\pi^{kl}}{\sqrt{\gamma}} - {\cal C}^{kl}) 
-{\cal K}^{-1}_{kl}{}^{mj} (D_j \An) (\frac{\pi^{kl}}{\sqrt{\gamma}} - {\cal C}^{kl}) \right] \cr
& + & 2 \alpha \sqrt{\gamma} [(D_m D_n N)-2(D_m N) D_n] \left[ \An {\cal K}^{-1}_{kl}{}^{mn} (\frac{\pi^{kl}}{\sqrt{\gamma}} - {\cal C}^{kl}) - D^n \tA^m \right]  
\eea
and
\bea\label{D3}
\frac{\delta {\cal H}_0(N)}{\delta \gamma_{ij}} & = & N [\cdots ] - (D^i N) \tA^j p_* + \sqrt{\gamma} [(D^i D^j N) - \gamma^{ij} (D^kD_k N) ] \\
& + & \alpha \sqrt{\gamma} (D_k N) \left[ \An (2  \tA^{(i} {\cal K}^{-1}_{mn}{}^{j)k} - \tA^k {\cal K}^{-1}_{mn}{}^{ij} )\frac{\pi^{mn}}{\sqrt{\gamma}}  
+ \sqrt{\gamma} (\tA^k D^i \tA^j - 2 \tA^{(i} D^{j)} \tA^k) \right] .\nonumber
\eea

We can now compute the various contributions to the Poisson bracket (\ref{HH}).
The part  which is linear in $p_\phi$ is by far the easiest to compute. It has a contribution from (\ref{D1}) only and is given by
\bea\label{p1}
\int d^3x [N_1(D^iN_2) - N_2 (D^i N_1)] \, A_i \, p_\phi \,.
\eea
The part linear in $p_*$ receives contributions from the three components of the Poisson bracket (\ref{HH}),
\bea
\int d^3x \frac{\delta {\cal H}_0(N_1)}{\delta \An } \frac{\delta {\cal H}_0(N_2)}{\delta p_*} & \rightarrow & 
\int d^3x \, 2{p_*} [N_1(D^iN_2) - N_2 (D^i N_1)] \left[ D_i \An - \tA^j {\cal K}^{-1}_{ij,kl} (\frac{\pi^{km}}{\sqrt{\gamma}} -{\cal C}^{kl})\right] \nonumber \\
\int d^3x  \frac{\delta {\cal H}_0(N_1)}{\delta \phi} \frac{\delta {\cal H}_0(N_2)}{\delta p_\phi} & \rightarrow & \nonumber
\int d^3x \,  [-N_1(D^iN_2) + N_2 (D^i N_1)]  \, p_* D_i \An \\
\int d^3x  \frac{\delta {\cal H}_0(N_1)}{\delta \gamma_{ij} } \frac{\delta {\cal H}_0(N_2)}{\delta \pi^{ij}} & \rightarrow & \nonumber 
\int d^3x \, 2{p_*} [N_1(D^iN_2) - N_2 (D^i N_1)] \tA^j {\cal K}^{-1}_{ij,kl} (\frac{\pi^{kl}}{\sqrt{\gamma}} -{\cal C}^{kl})\,,
\eea
which give the total contribution
\bea\label{p2}
\int d^3x \,  [N_1(D^iN_2) - N_2 (D^i N_1)]  \, p_* D_i \An \, .
\eea

The part linear in derivatives of $\pi^{ij}$ has contributions from the three components of (\ref{HH}) and is given by
\bea
[N_1 (D_m N_2)-  N_2 (D_m N_1)]\sqrt{\gamma}    \left[  4\alpha \tA^m A_i {\cal K}^{-1}_{kl}{}^{in}  - \An (1 +2\alpha \An + 2 \gamma^{mn} \gamma^{ij}) {\cal K}^{-1}_{kl}{}^{mn}   
 \right] D_n (\frac{\pi^{kl}}{\sqrt{\gamma}}) \nonumber
\eea
It is immediate the see that this expression reduces to
\bea
[N_1 (D_m N_2)-  N_2 (D_m N_1)]\sqrt{\gamma}  2 {\cal K}^{mn,ij} {\cal K}^{-1}_{ij,kl} D_n   (\frac{\pi^{kl}}{\sqrt{\gamma}}) 
\eea
which leads to
\bea\label{p3}
2 [N_1 (D_i N_2)-  N_2 (D_i N_1)] D_j   (\frac{\pi^{ij}}{\sqrt{\gamma}}) \, .
\eea

Gathering  (\ref{p1}), (\ref{p2}) and (\ref{p3}) and checking that the other contributions ( i.e. the terms quadratic in $\pi^{ij}$, those linear in $\pi^{ij}$ and those independent of  the momenta) vanish, we finally obtain
\bea
\{{\cal H}_0(N_1) , {\cal H}_0(N_2) \} = [N_1 (D_i N_2)-  N_2 (D_i N_1)]  {\cal H}^i\,.
\eea

\end{document}